\documentclass[aps,prd,nofootinbib]{revtex4-1}
\usepackage{graphicx}				
\usepackage{epsfig}
\usepackage{bm}
\usepackage{amssymb}
\usepackage{amsmath}
\usepackage{color}			
\usepackage{amssymb}

\usepackage{hyperref}
\hypersetup{
  colorlinks   = true,    
  urlcolor     = blue,    
  linkcolor    = blue,    
  citecolor    = red      
}

\newcommand{\be}{\begin{equation}}
\newcommand{\ee}{\end{equation}}
\newcommand{\bea}{\begin{eqnarray}}
\newcommand{\eea}{\end{eqnarray}}

\newcommand{\eq}[1]{Eq.~(\ref{#1})}

\newcommand{\fig}[1]{Fig.~\ref{#1}}

\newcommand{\dd}{{\rm d}}
\newcommand{\nn}{\nonumber}

\def\simge{\mathrel{%
   \rlap{\raise 0.511ex \hbox{$>$}}{\lower 0.511ex \hbox{$\sim$}}}}
\def\simle{\mathrel{
   \rlap{\raise 0.511ex \hbox{$<$}}{\lower 0.511ex \hbox{$\sim$}}}}

\begin{document}
\title{Multi-quark matrix elements in the proton and three gluon
  exchange for exclusive $\eta_c$ production in photon-proton diffractive
  scattering}

\author{Adrian Dumitru}
\email{adrian.dumitru@baruch.cuny.edu}
\affiliation{Department of Natural Sciences, Baruch College,
CUNY, 17 Lexington Avenue, New York, NY 10010, USA}
\affiliation{The Graduate School and University Center, The City
  University of New York, 365 Fifth Avenue, New York, NY 10016, USA}

\author{Tomasz Stebel}
\email{tomasz.stebel@ifj.edu.pl}
\affiliation{Institute of Nuclear Physics PAN, Radzikowskiego 152, 31-342 Krakow, Poland}
\affiliation{Physics Department, Brookhaven National Laboratory, Upton, NY 11973, USA}

\date{\today}

\begin{abstract}
Exclusive production of a $\eta_c$ pseudo-scalar meson in
$\gamma^{(*)} + p \to \eta_c + p$ scattering at high energies involves
a ${\cal C}$-odd exchange in the $t$-channel. We formulate the
description of this process within the high-energy framework of
eikonal dipole scattering. We obtain expressions for the light-cone
wave function of the $\eta_c$ required in this framework as well as
for the ${\cal C}$-odd amplitudes due to exchange of a single photon,
of a photon plus two gluons, and of three gluons. We relate these
amplitudes to correlators of the $+$ component of the quark current in
the light-cone wave function of the proton. For high transverse
momenta these correlators correspond to (leading twist) Generalized
Parton Distributions (GPDs) given by diagrams where all exchanged
gauge bosons attach to a single quark in the proton. Diagrams
involving multi-quark matrix elements potentially sensitive to
correlations, screen infrared singularities. Moreover, they are
numerically important for configurations where the exchanged bosons
nearly share the total momentum transfer.\\

Using two simple models for the three quark Fock state of the proton
at $x\simeq0.1$, we find that single photon exchange dominates for
$|t|<1.5$~GeV$^2$.  Here, the quark GPD could be measured cleanly in
$\gamma^{(*)} + p \to \eta_c + p$ via single photon exchange.  For
higher momentum transfer three gluon (``Odderon'') exchange is dominant.
\end{abstract}

\maketitle

\section{Introduction}

The existence of a color singlet three-gluon exchange with negative
${\cal C}$-parity in QCD at high energies was established long
ago~\cite{Bartels:1980pe}. Such an exchange could provide some
understanding of the difference of particle and anti-particle cross
sections and of the violation of the Pomeranchuk theorem in high-energy
scattering. An ``Odderon'' exchange has been proposed nearly 50 years
ago within the framework of Regge theory to explain the different
cross sections in ${\cal C}$-conjugate
channels~\cite{Lukaszuk:1973nt}. For a review of the theory and
experimental searches for the Odderon up until the year 2003 we refer
to ref.~\cite{Ewerz:2003xi}.

The TOTEM collaboration at the CERN-LHC has recently measured the
differential cross section for $pp$ elastic scattering at $\sqrt{s} =
2.76$~GeV~\cite{Antchev:2018rec}. They observe a significant
difference to the data by the D0 collaboration for $p\bar p$
scattering at $\sqrt{s} = 1.96$~GeV~\cite{Abazov:2012qb}. Assuming
that the difference in energy is negligible they conclude that these
results provide evidence for a color singlet 3-gluon exchange.
Even though these measurements are very exciting, the data does not
quite correspond to a kinematic regime where perturbative QCD may be
reliable.

Exclusive production of pseudo-scalar $\eta_c$ mesons in (virtual)
photon - proton scattering has been
highlighted~\cite{Schaefer,Czyzewski:1996bv,Engel:1997cga} as the
cleanest channel for discovery of ${\cal C}$-odd three gluon
(``Odderon'') exchange\footnote{Also see ref.~\cite{Kilian:1997ew} for
  a calculation of $\gamma\, p \to \eta_c\, p$ using Regge theory and
  effective Odderon-proton and Odderon-$\eta_c$ vertices. Moreover,
  one could search for Odderon exchange also in exclusive $\pi^+
  \pi^-$ pion pair electroproduction~\cite{Hagler:2002nf} or through
  exclusive vector meson production in proton-proton
  scattering~\cite{Bzdak:2007cz}.}. Here, the large mass of the
$c$-quark ensures that (at high energy) the process corresponds to
scattering of a small dipole of transverse extent much less than the
QCD color neutralization scale, from the proton. The focus in these
papers was on $\eta_c$ production at rather high energies and small
parton momentum fractions $x$, at HERA. However, the searches at HERA
did not observe exclusive $\eta_c$ production. The cross-section for
this process is small, in fact our estimates below are substantially
lower yet than old predictions from the
literature~\cite{Czyzewski:1996bv,Engel:1997cga}. Thus, such searches
for ${\cal C}$-odd three gluon exchange at a future high-luminosity
Electron Ion Collider (EIC) would be more promising\footnote{On the
  other hand, successful fits of exclusive $J/\Psi$ production at HERA
  energies (for example ref.~\cite{KMW}) can be interpreted to provide
  evidence for two-gluon Pomeron
  exchange~\cite{Low:1975sv,Nussinov:1975mw} supplemented by QCD
  high-energy evolution~\cite{Ryskin:1992ui}.}. A search for $\eta_c$
at Jefferson Lab may also be possible although our high-energy
scattering approximations would not provide accurate predictions for
near threshold energies. Their capabilities for measuring $J/\Psi$
production have been outlined in ref.~\cite{Joosten:2018gyo}.

Exclusive measurements in photon-proton scattering offer the
opportunity to extract fundamental nonperturbative QCD physics
contained in the light cone wave function of the
proton~\cite{Lepage:1980fj,Brodsky:1997de} and its Generalized Parton
Distributions
(GPDs)~\cite{Mueller:1998fv,Ji:1996ek,Radyushkin:1996nd,Collins:1996fb,Burkardt:2002hr,Diehl:2003ny,Belitsky:2005qn}.
Specifically, these processes in fact involve correlators of multiple
``+'' currents evaluated as matrix elements between multi-parton
states~\cite{DMV}. Here, we illustrate explicitly how they probe
multi-parton correlations in the proton.  We recover the description
in terms of a leading twist GPD when the transverse momenta of the
exchanged gluons or photon are large and generic.

We consider the following kinematic window in this paper. The
virtuality $Q$ of the photon is taken to be less than the mass $m_c$
of the charm quark, where the cross sections for $J/\Psi$ or $\eta_c$
production depend only weakly on $Q$. For higher photon virtuality the
$\eta_c$ cross section decreases further with increasing $Q$. We then
consider collision energies $W\sim 7 - 10$~GeV such that the process
probes quark momentum fractions in the proton of about $x\simeq0.1$;
c.f.\ eq.~(\ref{eq:x_vs_W}) below. At such $x$ it should be a
reasonable first approximation to describe the $\gamma^{(*)}\to
J/\Psi, \eta_c$ diffraction as eikonal dipole scattering. Indeed, the
longitudinal momentum transfer is much smaller than the transverse
momentum transfer considered here, c.f.\ eq.~(\ref{eq:tmin}) below.
At the same time, the proton state should still be dominated by its
valence quark Fock state. In the eikonal dipole scattering picture,
the dominant dipole size in $\eta_c$ production is about $r \sim
1$~GeV$^{-1}$ (slightly less for $J/\Psi$ production). At the same
time, we shall find that momentum transfers $|t|\simeq 1.5$~GeV$^2$ or
greater will be required to detect ${\cal C}$-odd three gluon
exchange in $\eta_c$ production. For such $r$ and $t$ an
interpretation in terms of gluon exchanges (between the quarks of the
proton and the $c\bar c$ dipole) appears reasonable.

\section{Setup}
The light cone state of an unpolarized on-shell proton with
four-momentum $P^\mu = (P^+, P^-,\vec{P}_\perp)$ is written
as~\cite{Lepage:1980fj,Brodsky:1997de}
\bea
|P\rangle &=& \frac{1}{\sqrt{6}} \int \frac{\dd x_1\dd x_2 \dd x_3}
{\sqrt{x_1 x_2 x_3}} 
\delta(1-x_1-x_2-x_3)
\int \frac{\dd^2 k_1 \dd^2 k_2 \dd^2 k_3}{(16\pi^3)^3}\,
 16\pi^3 \delta(\vec{k}_1+\vec{k}_2+\vec{k}_3)\nonumber\\
  &\times& 
 \psi_3(x_1, \vec k_1; x_2, \vec k_2; x_3, \vec k_3)
 \sum_{i_1, i_2, i_3}\epsilon_{i_1 i_2 i_3}
  |p_1,i_1, f_1; \, p_2,i_2, f_2; \, p_3,i_3, f_3\rangle~.  \label{eq:def_|P>}
\label{eq:valence-proton}
\eea
The $n$-parton Fock space amplitudes $\psi_n$ are universal and
process independent. They encode the non-perturbative structure of
hadrons.  Here, we have restricted ourselves to the valence quark Fock
state, assuming that the process probes parton momentum fractions of
order $x\sim 0.1$ or greater.  The three on-shell quark momenta are
specified by their lightcone momenta $p_i^+ = x_i P^+$ and their
transverse momenta $\vec{p}_{i} = x_i \vec{P}_\perp +
\vec{k}_i$. Colors and flavors of the quarks are denoted by
$i_{1,2,3}$ and $f_{1,2,3}$, respectively. In
eq.~(\ref{eq:valence-proton}) we have assumed that the wave function
$\psi_3$ is flavor blind, and we omit helicity quantum
numbers as they play no role in our analysis. $\psi_3$
is symmetric under exchange of any two of the quarks, and is
normalized according to
\be \label{eq:Norm_psi3}
 \int {\dd x_1\dd x_2 \dd x_3}\, \delta(1-x_1-x_2-x_3)
  \int \frac{{\dd^2 k_1 \dd^2 k_2 \dd^2 k_3}}{(16\pi^3)^3}\,
  (16\pi^3)\,\delta(\vec{k}_1+\vec{k}_2+\vec{k}_3)\, 
  |\psi_3|^2 = 1~.
\ee
This corresponds to the proton state normalization
\bea
\langle K | P\rangle &=& 16\pi^3 \, P^+ \delta(P^+ - K^+)
\, \delta(\vec{P}_\perp - \vec{K}_\perp) \label{eq:ProtonNorm1}
~.
\eea
Below, we neglect plus momentum transfer so that $\xi = (K^+ -
P^+)/P^+ \to 0$. This approximation is valid at high energies and $|t|
\gg |t_\text{min}|$. Accordingly, the light cone momentum of the
produced meson is close to that of the incoming photon.

For numerical estimates we shall employ the ``harmonic oscillator''
and ``power law'' model wave functions of Brodsky and
Schlumpf~\cite{Brodsky:1994fz},
\begin{eqnarray}
  \psi_{\rm H.O.}(x_1,\vec k_1; x_2, \vec k_2; x_3,\vec  k_3) &=& N_{\rm
    H.O.}\exp(-{\cal M}^2/2\beta^2)~, \nonumber\\
  \psi_{\rm Power}(x_1,\vec k_1; x_2, \vec k_2; x_3,\vec  k_3)
    &=& N_{\rm Power}(1+{\cal M}^2/\beta^2)^{-p}~.   \label{eq:pLFWF}
\end{eqnarray}
The invariant mass ${\cal M}$ of the configuration is given by
\begin{equation}
{\cal M}^2 = \sum_{i=1}^3 \frac{\vec k_{\perp i}^2+m^2}{x_i}~.
\end{equation}
$\beta$ determines the color neutralization scale and the typical
transverse momentum of quarks in the proton. The parameters $\beta$
and $m^2$ were obtained in ref.~\cite{Schlumpf:1992vq} by fitting to
electroweak parameters of the baryon octet: $m=0.26$ GeV, $\beta=0.55$
GeV for $\psi_{\rm H.O.}$ and $m=0.263$ GeV, $\beta=0.607$ GeV,
$p=3.5$ for $\psi_{\rm Power}$. The normalization constants $N_{\rm
  H.O.}$ and $N_{\rm Power}$ are obtained from the normalization
condition~(\ref{eq:Norm_psi3}).  Other models and parameter sets can
be found in
refs.~\cite{Frank:1995pv,Miller:2002ig,Pasquini:2007iz,Pasquini:2009bv,Lorce:2011dv}.\\

Following ref.~\cite{DMV} we introduce the charge density operators
corresponding to the light cone plus component of the quark currents
\bea
\rho(x_k,\vec k) &=& \sum_{f,i}
\int\frac{\dd x_q}{\sqrt{x_q(x_q+x_k)}}
 \int \frac{\dd^2q}{16\pi^3}\,
 b^\dagger_{q,i,f} b_{k+q,i,f} ~
 \substack{x_k\ll1\\ \longrightarrow}~
\sum_{f,i}
\int\frac{\dd x_q}{x_q}
 \int \frac{\dd^2q}{16\pi^3}\,
 b^\dagger_{x_q,\vec q,i,f} b_{x_q,\vec k+\vec q,i,f}
 \label{eq:rhoE} \\
\rho^a(x_k,\vec k) &=& \sum_{f,i,j}
\int\frac{\dd x_q}{\sqrt{x_q(x_q+x_k)}}
 \int \frac{\dd^2q}{16\pi^3}\,
 b^\dagger_{q,i,f} b_{k+q,j,f} \, (t^a)_{ij}~
  \substack{x_k\ll1\\ \longrightarrow}~
 \sum_{f,i,j}
\int\frac{\dd x_q}{x_q}
 \int \frac{\dd^2q}{16\pi^3}\,
 b^\dagger_{x_q,\vec q,i,f} b_{x_q,\vec k+\vec q,j,f} \, (t^a)_{ij} ~.
  \label{eq:rhoC} 
\eea
These equations define the densities of electric and color charge,
respectively; factors of $e$ and $g$ will be attached in
eqs.~(\ref{eq:A+_rho}) below. $b^\dagger_{q,i,f}$ and $b_{q,i,f}$
denote creation and annihilation operators for quarks with plus
momentum $q^+ = x_q P^+$, transverse momentum $\vec q$, color $i$, and
flavor $f$.

In what follows we shall neglect longitudinal momentum transfer to the
quarks and use the kinematic approximation where $x_k\sim 0.1 \ll1$.
This allows us to simplify the color charge operators as indicated
above. Moreover, we will assume that the scattering of a energetic
$c\bar c$ dipole from the valence charges in the proton is eikonal, to
first approximation; see eq.~(\ref{eq:Tamplitude_dipole_b})
below. Kinematic finite-$x$ corrections are suppressed by powers of
the light cone momentum $P^+$.  Of course, for quantitative
comparisons to future experiments it will be important to quantify these
corrections.

The charge densities are the sources for the static electromagnetic
and color fields in covariant gauge,
\be
\int \dd x^-\, A^+(x^-,\vec k) = \frac{e}{k^2}\, \rho(x_k= 0,
\vec k)~~~~~~,~~~~~~
\int \dd x^-\, A^{+a}(x^-,\vec k) = \frac{g}{k^2}\, \rho^a(x_k= 0,
\vec k)~. \label{eq:A+_rho}
\ee

The physical picture of representing the quarks with large light cone
momenta as static ($x^+$ independent) color charge densities sourcing
soft gluon fields was introduced by McLerran and
Venugopalan~\cite{MV}. In their effective theory, however, $\rho^a(x_k= 0,
\vec k)$ corresponds to a {\em classical} color charge vector
describing a large ensemble of quarks in a high-dimensional
representation of color-$SU(3)$. Here, instead, the operator $\rho^a(x_k= 0,
\vec k)$ acts on single quarks and color charge correlators will be
evaluated over the light cone wave function of the proton.

\section{Correlators of charge density operators in the proton}

In this section we provide expressions for correlators of various
(electric and color) charge density operators in the proton. Some of
these have been considered long before, especially for forward $K_T\to0$
scattering (see, for example ref.~\cite{Levin:1981rf}). Here we are
interested in non-forward matrix elements for single photon, two
gluon, photon plus two gluon, and three gluon exchanges. In
particular, we follow the approach of ref.~\cite{DMV} to relate these
matrix elements explicitly to the light cone wave function of the
proton, and to analyze their GPD limits.  \\

Consider first the expectation value of the electric charge density
operator in the proton, in the kinematic $x\ll1$ limit described
above. A straightforward calculation
yields\footnote{$\left<\cdots\right>_{\vec K_T}$ corresponds to
  $\langle K | \cdots | P\rangle$ stripped of the $\delta$-functions
  expressing conservation of transverse and plus momentum such as
  $\langle K |\rho(\vec q) | P\rangle = 16\pi^3 \, P^+
  \delta(P^+-K^+)\, \delta(\vec K_T + \vec q)\, \left<\rho(\vec
  q)\right>_{\vec K_T}$, if we set $\vec P_T=0$ for the incoming
  proton.}
\bea
\left< \rho(\vec q) \right>_{\vec K_T} &=& 
\sum_f e_f \, \int \dd x_1 \dd  x_2
\dd x_3 \, \delta(1-x_1-x_2-x_3) \nn\\
& &\times  \int \frac{\dd^2 p_1 \dd^2 p_2 \dd^2 p_3}{(16\pi^3)^2}
\, \delta(\vec{p}_1+\vec{p}_2+\vec{p}_3)\,
\psi_3^*(\vec p_1 + (1-x_1) \vec K_\perp, \vec p_2 -x_2 \vec K_\perp, \vec p_3
-x_3 \vec K_\perp)
\, \psi_3(\vec p_1, \vec p_2, \vec p_3) \nn\\
&\equiv& f(K_T)\, \sum_f e_f~.
\label{eq:VEV_rho_EM}
\eea
Here, $e_f$ denotes the fractional electric charge of quark
$f=(u,u,d)$, so that $\sum_f e_f=1$. For a lighter notation we omit
the arguments $x_1$, $x_2$ and $x_3$ of the wave functions here and
in similar expressions below.

The function $f(K_T)$ is a leading twist Generalized Parton
Distribution (GPD) of quarks in the proton at $x\ll1$, as it
corresponds to the non-forward matrix element of the plus component of
the quark current between single quark states. The photon probe can
only attach to one quark at a time and so this matrix element does not
probe correlations among the quarks. Also, for a single probe its
transverse momentum is equal to minus the recoil momentum of the
proton, $\vec q = - \vec K_T$.  For $K_T\to0$ the wave function
normalization in eq.~(\ref{eq:Norm_psi3}) implies $f(0)=1$.  \\

Next, we consider the correlator of two color charge density operators
which enters the amplitude for ${\cal C}$-even two-gluon exchange. It
is given by~\cite{DMV},
\bea
\langle \, \rho^a(\vec q) \, \rho^b(-\vec q-\vec K_T) \,\rangle_{\vec K_\perp}
&=&\,\frac{1}{2}\,\delta^{ab}\,\int \dd x_1 \dd  x_2
\dd x_3 \, \delta(1-x_1-x_2-x_3) \nn\\
& &\times  \int \frac{\dd^2 p_1 \dd^2 p_2 \dd^2 p_3}{(16\pi^3)^2}
\, \delta(\vec{p}_1+\vec{p}_2+\vec{p}_3)
\left[\psi_3^*(\vec p_1 + (1-x_1) \vec K_T, \vec p_2 -x_2 \vec K_T,
  \vec p_3 -x_3 \vec K_T) \right.\nn\\
  & & \left.
  -\psi_3^*(\vec p_1 +\vec q + (1-x_1)\vec K_T, \vec p_2 -\vec q -x_2
  \vec K_T, \vec p_3 -x_3 \vec K_T) \right]
\psi_3(\vec p_1, \vec p_2, \vec p_3) \label{eq:rho2_Kt_LFwf}\\
&\equiv& \frac{1}{2}\delta^{ab}\, G(\vec q, -\vec q - \vec K_T)~.
\label{eq:rho2_Kt}
\eea
$G(\vec q, -\vec q - \vec K_T)$ is real because the bound state wave
functions $\psi_3$ are real. It is invariant under a simultaneous
rotation of both $\vec q$ and $\vec K_T$ by the same angle and is also
symmetric under exchange of its two arguments, $G(\vec q, -\vec q -
\vec K_T) = G(-\vec q - \vec K_T, \vec q)$.  This color charge
correlator vanishes when $\vec q\to0$ or $\vec q_2 = -\vec q - \vec
K_T \to 0$ which expresses the color charge neutrality of the
proton. On the other hand, for large $|\vec q|$, $|\vec q_2|$, {\em
  and} large $|\vec q - \vec q_2|$, it is dominated by the first
term\footnote{For $\psi_3^* = \psi^*_\text{Power}$ and large momentum
  transfer $K_T$, it dominates by a power of $q^2 /K_T^2$ when
  $q^2\gg K_T^2$.}
in the square brackets of eq.~(\ref{eq:rho2_Kt_LFwf}). That is the
``one-body'' diagram where both gluon probes attach to the same
quark. In this GPD limit,
\be
G(\vec q, \vec q_2) \to
f(K_T)~~~~~~~~~(\text{for}~\vec q^{\,2}, \vec q^{\,2}_2,
(\vec q-\vec q_2)^2/4  \gg \Lambda^2_\text{eff})~.
\ee
Here, $\Lambda_\text{eff}\sim \beta$ is a soft scale encoded in the
light cone wave function of the proton, eqs.~(\ref{eq:pLFWF}). Note
that this approximation does not necessarily require large
$K_T$. However, at high momentum transfer, the contribution from the
diagram where the gluons couple to different quarks is important not
only for $\vec q \to 0$ or $\vec q_2 \to 0$ but also when $\vec q \sim
\vec q_2\sim -\vec K_T/2 \gg \Lambda_\text{eff}$. Here, the second
term in eq.~(\ref{eq:rho2_Kt_LFwf}) is actually much greater than the
first one due to contributions from configurations where the two
active quarks have large light cone momenta: at $K_T\gg
\Lambda_\text{eff}$ there is a large mismatch of the arguments of
$\psi_3^*$ and $\psi_3$ in the first term of
eq.~(\ref{eq:rho2_Kt_LFwf}) while there is strong overlap in the
second term when $\vec q \sim -\vec K_T/2$ and $x_1\sim x_2\sim
0.5$. We shall see below that restricting to the single quark matrix
element leads to a too rapid fall-off of $\dd\sigma^{J/\Psi}/\dd t$
with $|t|$.  \\

We now move on to the correlator of one electric charge with two color
charge densities which is relevant for ${\cal C}$-odd exchange of a
photon and two gluons. It is given by
\bea
\left< \rho(-\vec q_2-\vec q_3-\vec K_T) \, \rho^a(\vec q_2)\,
\rho^b(\vec q_3)\right>_{\vec K_T}
&=& 
\frac{1}{6}\delta^{ab}\, \sum_f e_f \, \int \dd x_1 \dd  x_2
\dd x_3 \, \delta(1-x_1-x_2-x_3)
\int \frac{\dd^2 p_1 \dd^2 p_2 \dd^2 p_3}{(16\pi^3)^2}
\, \delta(\vec{p}_1+\vec{p}_2+\vec{p}_3)\,\nn\\
& &\left[ 
\psi_3^*(\vec{p}_1+(1-x_1)\vec K_\perp ,\vec{p}_2-x_2\vec
K_\perp, \vec{p}_3-x_3\vec K_\perp) \right. \nonumber\\
& & +
2\psi_3^*(\vec{p}_1+\vec q_2+\vec q_3+(1 -x_1)\vec K_\perp ,
         \vec{p}_2-\vec q_2-\vec q_3 - x_2\vec K_\perp,
         \vec{p}_3-x_3\vec K_\perp) \nonumber\\
& &
  -
 \psi_3^*(\vec{p}_1+\vec q_2 +(1-x_1)\vec K_\perp ,\vec{p}_2-\vec q_2
 - x_2\vec K_\perp, \vec{p}_3-x_3\vec K_\perp) \nonumber\\
 & & -
 \psi_3^*(\vec{p}_1+\vec q_3 +(1-x_1) \vec K_\perp
  ,\vec{p}_2-\vec q_3- x_2\vec K_\perp,
 \vec{p}_3-x_3\vec K_\perp) \nonumber\\
 & & -
 \psi_3^*(\vec{p}_1+\vec q_2+\vec q_3 +(1-x_1)\vec K_\perp ,
 \vec{p}_2-\vec q_2 - x_2\vec K_\perp,
           \vec{p}_3- \vec q_3  -x_3 \vec K_\perp) \nn\\
  & & \Bigr]\, \psi_3(\vec p_1, \vec p_2 ,\vec p_3) \label{eq:VEV_rho_EM_rho2}\\
&\equiv& \frac{1}{2} \delta^{ab}\,
\left(\frac{1}{3}\sum_f e_f\right)\,
H(-\vec q_2- \vec q_3 -K_T, \vec
q_2, \vec q_3)~.
\label{eq:VEV_rho_EM_rho2_H}
\eea
This correlator vanishes when the transverse momentum of either one of
the attaching gluons vanishes, $\vec q_2=0$ or $\vec q_3=0$. When the
transverse momentum of the photon vanishes, $\vec q_1\equiv -\vec
q_2-\vec q_3-\vec K_T=0$ then it reduces to the correlator of two color
charge operators from eq.~(\ref{eq:rho2_Kt_LFwf}) times $\sum_f e_f =
1$.

For high transverse photon and gluon momenta, the first term in
eq.~(\ref{eq:VEV_rho_EM_rho2}) dominates and
\be
H(\vec q_1, \vec q_2, \vec q_3) \to
f(K_T)~~~~~~~~~(\text{for}~\vec q_1^{\,2}, \vec q_2^{\,2}, \vec q_3^{\,2},
(\vec q_i+\frac{1}{3}\vec K_T)^2 \gg \Lambda^2_\text{eff})~.
\ee
This is the leading twist GPD limit where both gluons and the photon
attach to the same quark in the proton. Note, however, that this
approximation does not apply when the exchanged bosons share the total
momentum transfer $\vec K_T$ in approximately equal proportions.  \\

Finally, we recall the correlator of three color charge densities
obtained in ref.~\cite{DMV},
\bea \langle \, \rho^a(\vec q_1) \, \rho^b(\vec q_2) \, \rho^c(-\vec
q_1-\vec q_2-\vec K_T) \,\rangle_{\vec K_\perp}
&=& \,\frac{1}{4}\,d^{abc}\,
\int \dd x_1 \dd x_2 \dd x_3 \,
\delta(1-x_1-x_2-x_3) \int \frac{\dd^2 p_1 \dd^2 p_2
  \dd^2 p_3}{(16\pi^3)^2} \, \delta(\vec{p}_1+\vec{p}_2+\vec{p}_3)
\nn\\
& &
\left[ 
\psi_3^*(\vec{p}_1+(1-x_1)\vec K_\perp ,\vec{p}_2-x_2\vec
K_\perp, \vec{p}_3-x_3\vec K_\perp) \right. \nonumber\\
& & -
  \psi_3^*(\vec{p}_1-\vec q_1 -x_1\vec K_\perp ,\vec{p}_2+\vec q_1
  +(1- x_2)\vec K_\perp, \vec{p}_3-x_3\vec K_\perp) \nonumber\\
  & &
  -
 \psi_3^*(\vec{p}_1+\vec q_2 +(1-x_1)\vec K_\perp ,\vec{p}_2-\vec q_2
  - x_2\vec K_\perp, \vec{p}_3-x_3\vec K_\perp) \nonumber\\ & & -
 \psi_3^*(\vec{p}_1-\vec q_1-\vec q_2 -x_1 \vec K_\perp
  ,\vec{p}_2+\vec q_1 +\vec q_2+ (1- x_2)\vec K_\perp,
  \vec{p}_3-x_3\vec K_\perp) \nonumber\\ & & +2\, 
\psi_3^*(\vec{p}_1-\vec
  q_1 -x_1\vec K_\perp ,\vec{p}_2+\vec q_1 +\vec q_2+  (1-x_2) \vec K_\perp,
  \vec{p}_3-\vec q_2 - x_3\vec K_\perp ) \nn\\
  & & \Bigr]\, \psi_3(\vec p_1, \vec p_2 ,\vec p_3) \label{eq:rho^3correl}\\
&\equiv&
\frac{1}{4}d^{abc}\, G_O(\vec q_1, \vec q_2, -\vec q_1 - \vec q_2
- \vec K_T)~.
\label{eq:rho3_Kt}
\eea
(We have redefined the normalization of $G_O$ as compared to
ref.~\cite{DMV} by a factor of $N_c/4$ for convenience.)  Here, on the
r.h.s.\ we have written only the $C$-odd contribution which is
symmetric under the exchange of the color indices $a$, $b$, and $c$
since this is the only piece that couples to a dipole. $G_O$ is
invariant under a simultaneous rotation of the three transverse
momentum vectors $\vec q_1$, $\vec q_2$, $\vec K_T$ and under
permutations of its arguments. When either one of the three momenta
$\vec q_1$, $\vec q_2$, or $\vec q_3\equiv -\vec q_1 - \vec q_2 - \vec
K_T$ is zero then $G_O$ vanishes. Once again one recovers the leading
twist GPD corresponding to the matrix element of the plus current
between single quark states in the limit of large transverse momenta
of the gluon probes,
\be
G_O(\vec q_1, \vec q_2, \vec q_3) \to
f(K_T)~~~~~~~~~(\text{for}~\vec q_1^{\,2}, \vec q_2^{\,2}, \vec q_3^{\,2},
(\vec q_i+\frac{1}{3}\vec K_T)^2  \gg \Lambda^2_\text{eff})~.
\ee
This approximation again requires not only large transverse gluon
momenta but also that the total momentum transfer is not being
(approximately) shared equally.\\

In closing this section we note that in our expressions for the color
charge correlators $\langle \, \rho^a(\vec q) \, \rho^b(-\vec q-\vec
K_T) \,\rangle_{\vec K_\perp}$ etc.\ we have ignored the appropriate
gauge links connecting the
sources~\cite{Dominguez:2011wm,Altinoluk:2019fui,Boer:2018vdi}. These
gauge links account for soft multiple scattering while we retain only
two or three ``hard'' gluon exchanges with the target. In this paper
we shall be interested mainly in the limit $K_T^2 \gg \Lambda^2_\text{eff}$.

\section{Dipole scattering amplitude}

\begin{figure}[htb]
  \centering
  \includegraphics[width=0.45\textwidth]{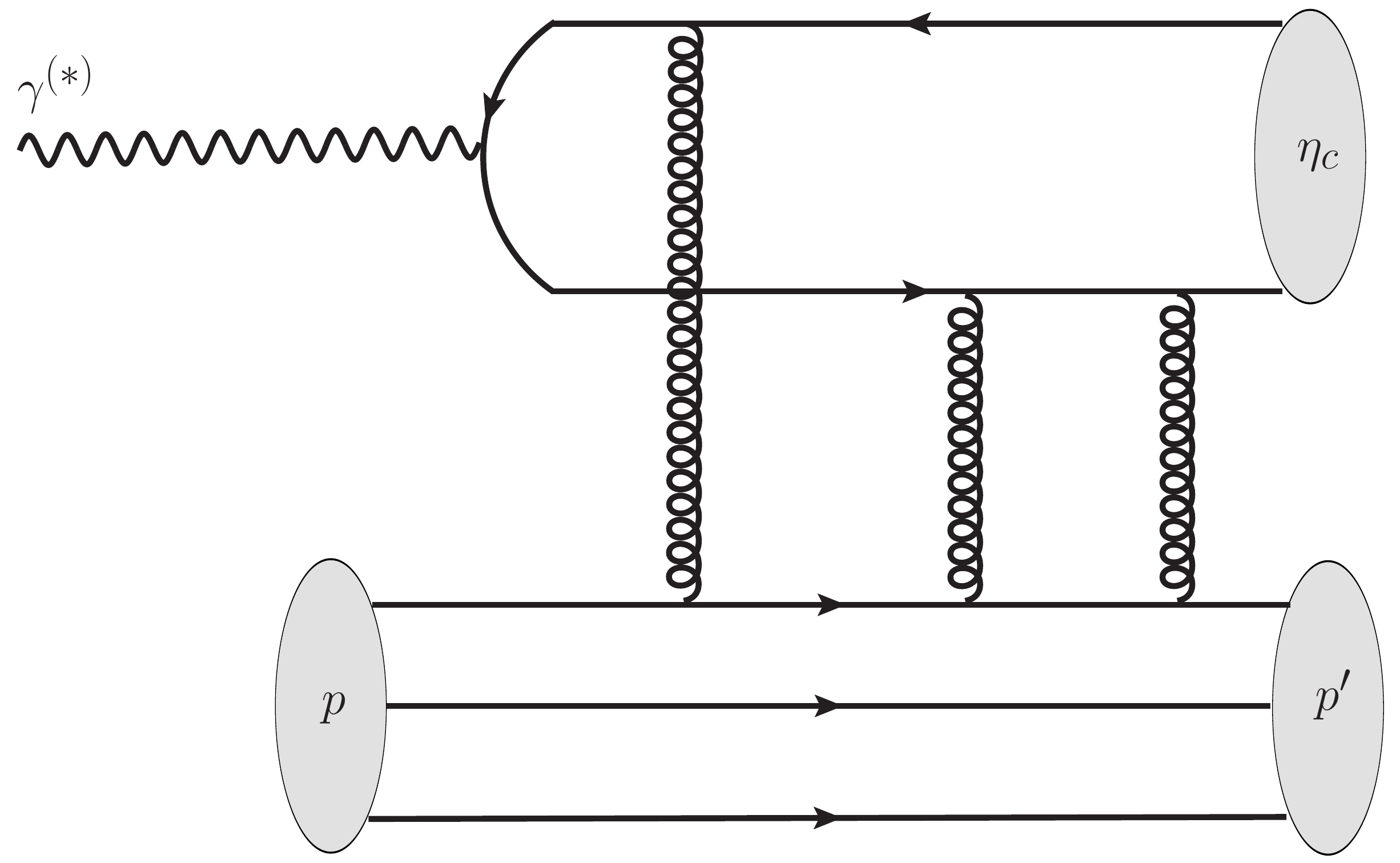}
  \includegraphics[width=0.45\textwidth]{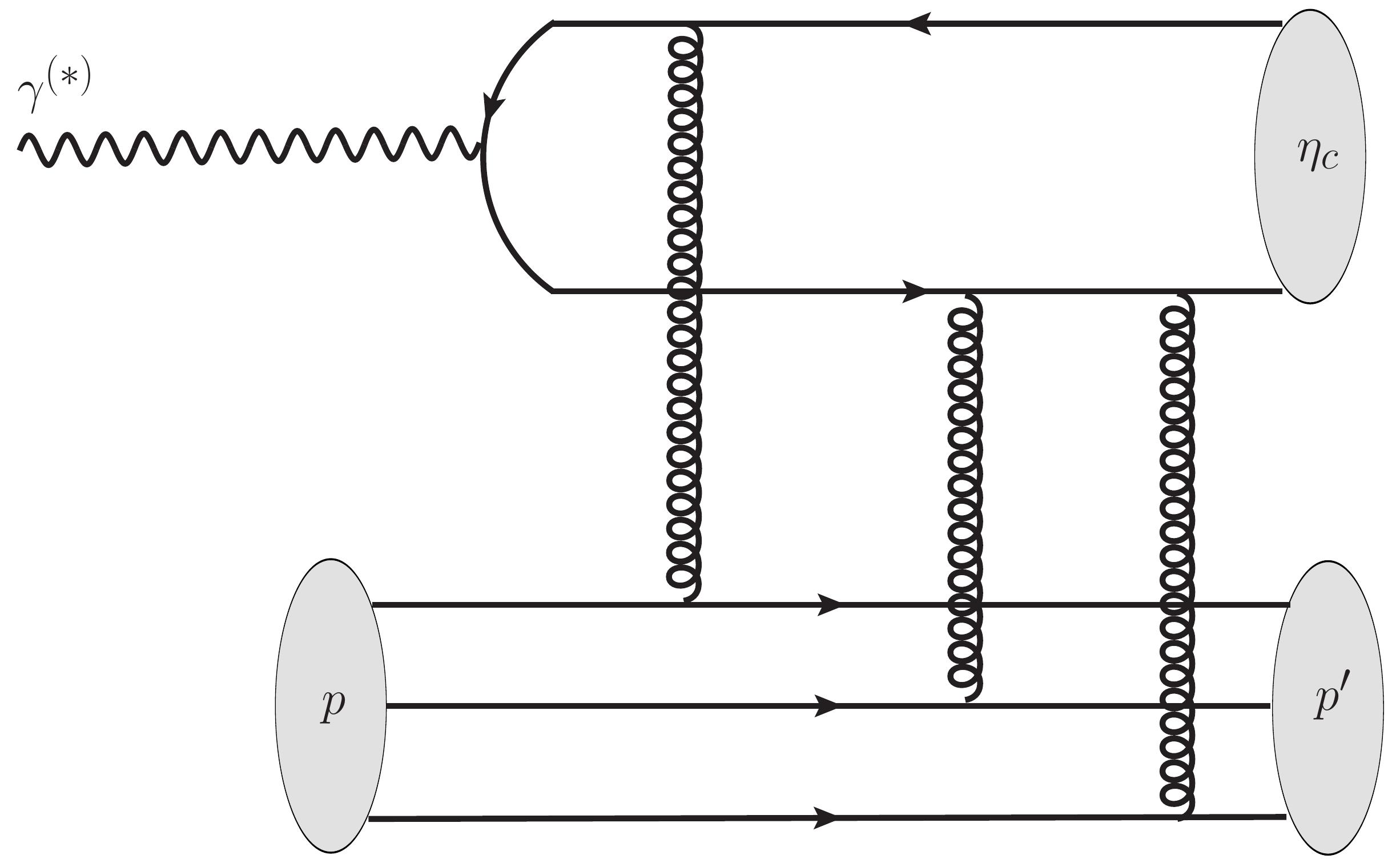}
  \caption{Two of the diagrams that contribute to the production of a
    pseudoscalar $\eta_c$ meson via ${\cal C}$-odd 3-gluon
    exchange. The diagram on the left involves a matrix element
    between single quark states in the proton while the diagram on the
  right involves a matrix element in a three quark state sensitive to
  multi-quark correlations.}
\label{fig:3gXchange}
\end{figure}

The invariant amplitude ${\cal T}$ for elastic scattering of the
$c{\bar c}$ pair off the fields in the target proton can be
expressed as
\bea \label{eq:Tamplitude_dipole_b}
   {\cal T} (\vec r,\vec b_\perp; \vec K_\perp) &=& 2\, N_c\,
   \left[ 1 - \frac{1}{N_c}\,{\rm tr} \,\left< U\left(\vec b +
     \frac{\vec r}{2}\right)
     U^\dagger\left( \vec b - \frac{\vec r}{2}\right)\right>_{\vec
       K_\perp} \right]\, , \label{eq:Tij} \\
   {\cal T} (\vec r, \vec K_\perp) &=& \int \dd^2b \, e^{i \vec b \cdot
     \vec K_T} \, {\cal T} (\vec r,\vec b_\perp; \vec K_\perp)~.
    \label{eq:Tamplitude_dipole_Kt}
\eea
At $K_T=0$ eq.~(\ref{eq:Tij}) is related to the so-called dipole gluon
distribution evaluated in covariant gauge~\cite{Dominguez:2011wm}.
Here, $U$ (and $U^\dagger$) are lightlike Wilson lines representing the
eikonal scattering of the dipole of size $\vec r$ at impact parameter
$\vec b$. Two of the diagrams that contribute to the ${\cal C}$-odd
part of this amplitude are shown in fig.~\ref{fig:3gXchange}.

To account for photon exchange to the scattering
amplitude~(\ref{eq:Tamplitude_dipole_b}) we use Wilson lines in the
combined color and electromagnetic fields:
\be
U^\dagger(\vec x_T) = {\cal P} e^{i \int dx^- \left[gA^{+a}(x^-,\vec
    x_T)\, t^a + e_Q e A^+(x^-,\vec x_T)\right]} ~.
\ee
Here, $e_Q=2/3$ is the fractional charge of the $c$-quark.

In what follows we expand ${\cal T} (\vec r, \vec K_\perp)$ up to
first order in the electromagnetic field, and up to third order in the
color field. The relation of such a weak field expansion to a
resummation of kinematic twists in Wandzura-Wilczek type
approximations has been elucidated recently in
ref.~\cite{Altinoluk:2019fui}. Indeed, we do not expand the scattering
amplitude about small dipole size $r$ or small momentum exchange
$K_T$.  However, as already indicated at the end of the previous
section, we do neglect the resummation of multiple soft
scattering. For a proton target and $x\sim 0.1$ the weak field limit
should provide a reasonable first approximation.  \\

Expanding to first order in the fields we obtain the amplitude for
single photon exchange,
\be
{\cal T}_\gamma (\vec r, \vec K_\perp) = 16\pi N_c\, \alpha\, e_Q\,
\sum_f e_f\, \frac{f(K_T)} {K_T^2}\, \sin\left( \frac{\vec r\cdot \vec
  K_T}{2}\right)~,
\ee
with $f(K_T)$ from eq.~(\ref{eq:VEV_rho_EM}). Note that the exchanged
photon is off shell as it possesses only transverse but no light cone
momentum.\\

The contribution at second order in the color field $gA^{+a}$
corresponds to ${\cal C}$-even two gluon (``Pomeron'') exchange~\cite{DMV},
\bea
{\cal T}_{gg}(\vec r,\vec K_\perp) &=& 
  -(4\pi\alpha_s)^2 N_c C_F\,\int\limits_q
  \frac{1}{(\vec q -\frac{1}{2}\vec K_T)^2 \, (\vec q+ \frac{1}{2} \vec
  K_T)^2}\,\left(
\cos\left(\vec r \cdot {\vec q}\right) 
- \cos\left(\frac{{\vec r}\cdot \vec K_T}{2}\right)\right) \nn\\
& & \hspace{4cm} \times \,
G\left({\vec q}-\frac{1}{2}\vec K_T,-{\vec q}-\frac{1}{2}\vec K_T\right)
~.    \label{eq:Pomeron}
\eea
(We use the shorthand notation $\int_q = \int\dd^2q/(2\pi)^2$.)
${\cal T}_{gg}(\vec r,\vec K_\perp)$ is even under a sign flip of either
$\vec r$ or $\vec K_T$. For forward scattering of a small dipole,
\be
{\cal T}_{gg}(\vec r, K_\perp=0) \simeq \frac{1}{4}
(4\pi\alpha_s)^2 N_c C_F \, r^2\log\frac{1}{r\, \Lambda_\text{eff}}~,
~~~~~~~~~~(\text{for}~r\Lambda_\text{eff}\ll1)~,
\ee
exhibits the well-known ``color transparency'' effect. The logarithm
in the previous expression arises because the transverse momenta of
the exchanged gluons are distributed from $\Lambda_\text{eff}$ to the
hard scale $r^{-1}$ according to $\dd q^2/q^2$.\\

At second order in $gA^{+a}$ and first order in $eA^+$ we have
\bea
{\cal T}_{\gamma gg}(\vec r, \vec K_\perp) &=& \frac{1}{2} e_Q \,
4\pi \alpha\, (4\pi\alpha_s)^2 \, (N_c^2-1)\, \left(\frac{1}{3}\sum_f
e_f\right) \int\limits_{q_2,q_3}
\frac{1}{q_2^2}\frac{1}{q_3^2} \frac{1}{(\vec q_2+\vec q_3+\vec
  K_T)^2} \left[ \sin\left(\vec q_2\cdot \vec r+\frac{1}{2}\vec
  K_T\cdot r\right) \right.\nn\\
  & & ~~~~~~~~~~~~~~\left.  +
  \sin\left(\vec q_3\cdot \vec r+\frac{1}{2}\vec K_T\cdot r\right)
  -\sin\left(\frac{1}{2}\vec K_T\cdot r\right) - \sin\left((\vec
  q_2+\vec q_3)\cdot \vec r+\frac{1}{2}\vec K_T\cdot r\right)
  \right]\nn\\ & & ~~~~~~~~~~\times \, H(-\vec q_2-\vec q_3-\vec K_T,
\vec q_2, \vec q_3)~,
\eea
with $H$ as given in eq.~(\ref{eq:VEV_rho_EM_rho2}).  The integrand
does not exhibit any infrared divergences at $\vec q_2=0$ or $\vec
q_3=0$ or $\vec q_2+\vec q_3+\vec K_T=0$.  ${\cal T}_{\gamma gg}(\vec
r, \vec K_\perp)$ is odd under a sign flip of either $\vec r$ or $\vec
K_T$.\\

Finally, at third order in $gA^{+a}$ we have the following scattering
amplitude for ${\cal C}$-odd three gluon
exchange\footnote{Ref.~\cite{DMV} denotes this amplitude $iO(\vec
  r,\vec K_\perp)$. Our expression in eq.~(\ref{eq:Odderon-operator})
  includes a factor of $-2N_c$ omitted in
  ref.~\cite{DMV} in the step from their eqs.~(74) to~(76).}~\cite{DMV}:
\bea
{\cal T}_{ggg}(\vec r,\vec K_\perp) &=&  
\frac{5}{3}\, (4\pi\alpha_s)^3 
\int\limits_{q_1, q_2}
\frac{1}{q_1^2}\frac{1}{q_2^2}\frac{1}{q_3^2}\,
G_O(\vec q_1,\vec q_2,\vec q_3)\,\left[
\sin\left(\vec r\cdot \vec q_1 + \frac{1}{2} \vec r\cdot \vec
    K_T\right)
    - \frac{1}{3}\sin\left(\frac{{1}}{2}\vec r\cdot
  \vec K_T\right)\right]~.
\label{eq:Odderon-operator}
\eea
Here, $\vec q_3 \equiv -\vec q_1 - \vec q_2 - \vec K_T$.  ${\cal
  T}_{ggg}(\vec r, \vec K_\perp)$ is also odd under a sign flip of
either $\vec r$ or $\vec K_T$. The relation of the Odderon amplitude
to the $T$-odd gluon GTMDs and GPDs in the gluon dominated regime of
very small $x$ has been worked out in ref.~\cite{Boer:2018vdi}.

In the limit of nearly forward scattering of a small dipole the Odderon
behaves differently than the Pomeron. The second term in the equation
above, for example, is
\bea
\sim
\sin\left(\frac{{1}}{2}\vec r\cdot \vec K_T\right)
\int\limits_{q_1}\int\limits_{q_2}
\frac{1}{q_1^2}\frac{1}{q_2^2}\frac{1}{q_3^2}\, G_O(\vec q_1,\vec q_2,\vec q_3)
~.
\label{eq:Odderon-b}
\eea
The integral in this expression is independent of the hard scale $1/r$
set by the size of the dipole. For nearly forward scattering all three
exchanged gluons will have transverse momenta of order of the soft
color neutralization scale $\Lambda_\text{eff}$ because the integrands
drop faster than $\dd^2 q_i/q_i^2$. On the other hand, if one requires
a large momentum transfer $K_T^2$ then $q_2$ in
eq.~(\ref{eq:Odderon-operator}) runs from $\Lambda_\text{eff}$ to
$K_T$ while $q_1$ extends from $\Lambda_\text{eff}$ to
$\text{min}(K_T,1/r)$. Hence, for large $K_T$ but small $r$, with
$\vec r\cdot \vec K_T\sim1$ or greater, and neglecting the
contribution from $\vec q_1 \sim \vec q_2 \sim \vec q_3 \sim - \vec
K_T/3$, the leading logarithmic contribution to ${\cal T}_{ggg}$ is
\be
{\cal T}_{ggg}(\vec r,\vec K_\perp) \simeq
\frac{10}{9}\, (4\pi\alpha_s)^3 
\sin\left(\frac{{1}}{2}\vec r\cdot \vec K_T\right)
f(K_T)\,  \frac{\log(K_T/\Lambda_\text{eff})\,
  \log(1/r\Lambda_\text{eff})}{(2\pi)^2\, K_T^2}~.
\ee
This expression again involves the same GPD $f(K_T)$ encountered
above. 
\\

We have indicated in the previous section that the correlators of
multiple charge density operators in the proton reduce to the
expectation value of a {\em single} such operator (a leading twist
GPD) when the transverse momenta of the attached gauge bosons are
large and not close to each other. In that regime, the correlators are
dominated by the diagram where all gauge bosons couple to the same
quark.  We shall illustrate the importance of the diagrams involving
multi-quark matrix elements as follows\footnote{As already mentioned,
  here we restrict to the valence quark component of the proton wave
  function.  For a discussion of correlations among quarks in the
  proton at $x\ll0.1$ see, for example,
  ref.~\cite{Altinoluk:2016vax}.}. In eqs.~(\ref{eq:rho2_Kt_LFwf})
and~(\ref{eq:rho^3correl}) for the two and three gluon exchange matrix
elements\footnote{We do not discuss ${\cal T}_{\gamma gg}$ in this
  context as its contribution turns out to be very small.},
respectively, we drop all but the first contribution in the square
brackets; which is the diagram where all exchanged gluons couple to a
single quark. The respective correlators at fixed total momentum
transfer $\vec K_T$ are then independent of the transverse momenta of
the attached gluons.  Hence, $G\left({\vec q}-\frac{1}{2}\vec
K_T,-{\vec q}-\frac{1}{2}\vec K_T\right) \to f(K_T)$ in
eq.~(\ref{eq:Pomeron}), and $G_O(\vec q_1,\vec q_2,\vec q_3) \to
f(K_T)$ in eq.~(\ref{eq:Odderon-operator}).  However, now the
contributions to the integrals in eqs.~(\ref{eq:Pomeron},
\ref{eq:Odderon-operator}) are no longer cut off at soft transverse
momentum.  To restore color screening at low transverse momentum we
therefore introduce by hand cutoffs of the form
\be
\left( 1- e^{-\frac{(\vec q-\vec K_T/2)^2}{2\Lambda^2}}\right)
\left( 1- e^{-\frac{(\vec q+\vec K_T/2)^2}{2\Lambda^2}}\right)
~~~~~~,~~~~~
\left( 1- e^{-\frac{q_1^2}{2\Lambda^2}}\right)
\left( 1- e^{-\frac{q_2^2}{2\Lambda^2}}\right)
\left( 1- e^{-\frac{q_3^2}{2\Lambda^2}}\right)
\label{eq:cutoff}
\ee
in eqs.~(\ref{eq:Pomeron}, \ref{eq:Odderon-operator}). In particular,
for three gluon exchange such ad hoc cutoffs are unavoidable if one
restricts to the one-body diagram as ${\cal T}_{ggg}$ is otherwise
infrared divergent.

We shall use $\Lambda=0.1$~GeV for numerical estimates. We have not
attempted to ``fine tune'' the cutoff $\Lambda$ to the color
neutralization scale encoded in the light cone wave function of the
proton. However, we have checked that imposing such a cutoff on the
complete set of diagrams does not affect the cross sections
much. Below we show the numerical accuracy of these ``single quark +
cutoff'' approximations for the $J/\Psi$ and $\eta_c$ cross sections.

\section{Exclusive $J/\Psi$ and $\eta_c$ production in
  $\gamma^{(*)}-p$ scattering}

\subsection{Light cone wave function of the $\eta_c$}

Before discussing the amplitude for exclusive meson production in
$\gamma^*p\rightarrow M p$ we need to derive the light cone wave
function of the $\eta_c$ required in the dipole scattering approach.\\

We take the spinor part as $i\gamma^5$ sandwiched between
$\bar{u}_h (z,\vec k_T)$ and $v_{\bar h} (1-z,-\vec k_T)$ spinors, in
the transverse rest frame of the $\eta_c$. This is multiplied by a
phenomenological scalar wave function $\phi^P(k_T,z)$ like for the
$J/\Psi$ meson, c.f.\ appendix~\ref{sec:LCwf}:
\be
\Psi^{\eta_c}_{h,\bar h}(\vec{k}_T,z)=-i\frac{\bar{u}_h (z,\vec k_T)}{\sqrt{z}}
\gamma^5\, \frac{v_{\bar h} (1-z,-\vec k_T)}{\sqrt{1-z}}\, \tilde
\phi^{P}(k_T,z).
\label{eta_wf_kTspace}
\ee
Here, $z$, $\vec k_T$ and $h$ denote the light cone momentum fraction,
the transverse momentum, and the helicity of the
$c$ quark; $1-z$, $-\vec k_T$ and $\bar h$ those of the $\bar c$ anti-quark.

The spinor matrix element can be computed using the expressions
summarized in refs.~\cite{Li_spinors,Lepage:1980fj}:
\bea
-\bar{u}_h(z, \vec k_T) \gamma^5 v_{\bar h} (1-z,-\vec k_T) &=& 
\frac{1}{\sqrt{z(1-z)}} \left[-k^R_T \, \delta_{h-}\delta_{\bar h-}
- k^L_T\, \delta_{h+}\delta_{\bar h+}
+m_c \left( \delta_{h+}\delta_{\bar h-} -  \delta_{h-}\delta_{\bar h+}\right)
\right]~.
\eea
$m_c$ denotes the mass of the charm quark which we take as 1.4~GeV~\cite{KMW}.
Here, the transverse momenta are written in complex representation as
$k^{R,L}_T = k^1 \pm i k^2$. Written as operators in coordinate
representation, $k^{R,L}_T \to \pm i \, e^{\pm i \phi_r}\,
\partial_r$. Then
\be
\Psi^{\eta_c\ }_{h,\bar h}(\vec{r},z) = \frac{i}{z(1-z)} \left[ i
  \left( \delta_{h+,\bar h+} e^{-i \phi_r}- \delta_{h-,\bar h-} e^{i
    \phi_r} \right) \partial_r - m_c\left( \delta_{h-,\bar h+} -
  \delta_{h+,\bar h-} \right) \right] \phi^P(r,z)~.
\label{eta_wf_rspace}
\ee
The scalar part $\phi^P(r,z)$ of the pseudoscalar meson wavefunction
has to be modeled.  In order to fix $\phi^P(r,z)$ we adopt a simple
approach and use the same ``Boosted Gaussian'' functional form of the
scalar function as for the vector
meson~\cite{KMW,Forshaw:2003ki,Nemchik:1994fp,Nemchik:1996cw}:
\begin{align}
  \phi^{P}(r,z) = \mathcal{N}_{P}\, z(1-z)
  \exp\left(-\frac{m_c^2 \mathcal{R}_P^2}{8z(1-z)} -
  \frac{2z(1-z)r^2}{\mathcal{R}_P^2} +
  \frac{m_c^2\mathcal{R}_P^2}{2}\right)~.
  \label{Boosted-Gaussian_wf_model_P}
\end{align}

To fix the parameters $\mathcal{R}^2_P$ and $\mathcal{N}_P$ we impose
the normalization condition
\begin{equation}
  1 = N_c \sum_{h,\bar h} \int \dd^2 r 
  \int_0 ^1 \frac{d z}{4\pi}\, \left\lvert\Psi^{\eta_c} _{h\bar
    h}(\vec{r},z,Q^2)\right\rvert^2,
  \label{fullnorm_1}
\end{equation}
which, after substitution of~(\ref{eta_wf_rspace}) becomes:
\begin{equation}
1 = \frac{N_c}{2\pi} \sum_{h,\bar h} \int \dd^2 r 
\int_0 ^1 \frac{d z}{z^2(1-z)^2}\, \left[ \left( \partial_r
  \phi^P \right)^2 +  \left( m_c \phi^P \right)^2      \right]~.
  \label{fullnorm_2}
\end{equation}
We have assumed that $\phi^P(r,z)$ is real.

A second constraint on the wave function arises from the requirement
that it matches the coupling to the axial-vector current,
\be \label{eq:f_P}
\left<0\left| \bar c(0) \gamma^\mu \gamma_5 c(0) \right| P\right> = i f_P P^\mu~.
\ee
The meson state with momentum $P^\mu$ is written as
\be \big| P\bigr> = \frac{\widetilde{N}_P}{\sqrt{N_c}} \sum_{h,\bar h,
  i} \int \frac{\dd x \, \dd^2k_T}{16\pi^3\, \sqrt{x (1-x)}}
\,\Psi^{\eta_c}_{h\bar h}(x,\vec k_T) \, \Big| x,\vec k_T - x\vec P_T,
h, i;\, 1-x,- \vec k_T - (1-x)\vec P_T, \bar h, i \Bigr>~.
\ee
Here, $i=r,g,b$ denotes color and $h$, $\bar h$ are the helicities of
the $c$-quarks. The normalization condition~(\ref{fullnorm_1})
requires $\widetilde{N}_P=\sqrt{N_c}$ in order to make sure that
\be
\left< K \big| P\right> = 16\pi^3\, P^+ \delta(P^+-K^+)\, \delta(\vec
P_T - \vec K_T)~.
\ee
We now introduce quark and anti-quark creation and annihilation
operators through the bare field expansions~\cite{Lepage:1980fj}
\bea
c_i(x^\mu) &=& \int \frac{\dd x_p \, \dd^2 p_T}{16\pi^3 x_p} \,
\sum_h \left[b_{hi}(p)\, u_h(p)\, e^{-i  p\cdot r}
  + d^\dagger_{hi}(p)\, v_h(p)\, e^{i  p\cdot r}\right]~, \\
\bar c_i(x^\mu) &=& \int \frac{\dd x_p \, \dd^2 p_T}{16\pi^3 x_p} \,
\sum_h \left[b^\dagger_{hi}(p)\, \bar u_h(p)\, e^{i  p\cdot r}
  + d_{hi}(p)\, \bar v_h(p)\, e^{-i  p\cdot r}\right]~.
\eea
These satisfy the anti-commutation relations
\be
\left\{b_{hi}(p), \, b^\dagger_{h'i'}(k) \right\}
=\left\{d_{hi}(p), \, d^\dagger_{h'i'}(k) \right\} = 16\pi^3\,
p^+\delta(p^+-k^+) \, \delta(\vec p_T - \vec k_T)\, \delta_{ii'}\,\delta_{hh'}~.
\ee

A straightforward calculation of the $+$ component of
eq.~(\ref{eq:f_P}), using again the spinor matrix elements summarized
in ref.~\cite{Li_spinors}, now leads to
\bea
f_P &=& \widetilde{N}_P\, \sqrt{N_c} \int \frac{dz}{\pi} \, \frac{m_c}{z (1-z)}
\, \Phi^P(z, r=0)
\label{eq:f_P_constraint}
\eea
The leading order decay rate of the $\eta_c$ to two photons is related
to $f_P$ through (see, for example, ref.~\cite{Pham:2007xx})
\be \label{eq:Gamma2photon}
\Gamma_{\eta_c\to \gamma \gamma} = 4\pi e_c^4 \alpha^2\frac{f_P^2}{M_{\eta_c}}~.
\ee
Experimentally~\cite{PDG},
\be  \label{eq:Meta_fP}
M_{\eta_c} = 2.984~\mathrm{GeV}~~~~,~~~~
\Gamma_{\eta_c\to \gamma \gamma} = 5.04\times10^{-6}~\mathrm{GeV}~.
\ee
Eqs.~(\ref{eq:Gamma2photon}, \ref{eq:Meta_fP}) give
$f_P=0.337$~GeV. We employ $m_c=1.4$~GeV both for the $\eta_c$ and
$J/\Psi$ wave functions. The two constraints~(\ref{fullnorm_1})
and~(\ref{eq:f_P_constraint}) determine the two parameters of the
$\eta_c$ wave function to be $\mathcal{N}_{P}=0.547$ and
$\mathcal{R}_{P}=2.48$~GeV$^{-2}$. As expected, these values are close
to those for the $J/\Psi$ meson given in the appendix. \\

The products of photon and $\eta_c$ wave functions are
\bea
\left[\left(\Psi^{\eta_c}\right)^*\,
  \Psi^{\gamma^*}_L\right]_{{\substack{h=\pm\\ \bar
      h=\mp}}}(\vec{r},z,Q^2) &=&  i\,\mathrm{sign}(h)\,
\frac{m_c \, e_c \,e \,Q}{\pi} \,
K_0(\epsilon r) \, \phi^P(z,r) ~, \label{eq:PsiPsi_L_hhbar}\\
\left[\left(\Psi^{\eta_c}\right)^*\,
  \Psi^{\gamma^*}_{T,\lambda=\pm}\right]_{h\bar h}(\vec{r},z,Q^2) &=&
- \frac{\sqrt{2}\, e_c e\,  }{2\pi\,  z(1-z)} \, m_c \, e^{ i\lambda\phi_r}\nn\\
& & \hspace{-1.5cm}\times\left\{
K_0(\epsilon r) \partial_r \phi^P\delta_{\lambda h=+}\delta_{\lambda \bar h=+}
+ \epsilon K_1 (\epsilon r) \phi^P \left[
  z\delta_{\lambda h=+}\delta_{\lambda \bar h=-}
  + (1-z) \delta_{\lambda h=-}\delta_{\lambda \bar h=+}
  \right]\right\}~,
\eea
where $\epsilon \equiv \sqrt{z(1-z)Q^2+m_c^2}$.
Summing the amplitude over quark helicities gives
\bea
\left[\left(\Psi^{\eta_c}\right)^*\,
  \Psi^{\gamma^*}_L\right](\vec{r},z,Q^2) &=&  0 ~, \label{eq:PsiPsi_L}\\
\left[\left(\Psi^{\eta_c}\right)^*\,
  \Psi^{\gamma^*}_{T,\lambda=\pm}\right](\vec{r},z,Q^2) &=&
- \frac{\sqrt{2}\, e_c e\,  }{2\pi\,  z(1-z)} \, m_c \, e^{ i\lambda\phi_r}
\left\{
K_0(\epsilon r) \partial_r \phi^P
+ \epsilon K_1 (\epsilon r) \phi^P\right\}~.
\eea
The transverse amplitude corresponds to an average over $\lambda=\pm$,
\be
\left[\left(\Psi^{\eta_c}\right)^*\,\Psi^{\gamma^*}_T\right](\vec{r},z,Q^2) =
- \frac{\sqrt{2}\, e_c e\,  }{2\pi\,  z(1-z)} \, m_c \, \cos(\phi_r)
\left\{
K_0(\epsilon r) \partial_r \phi^P
+ \epsilon K_1 (\epsilon r) \phi^P\right\}~.
\ee

\subsection{Amplitudes and cross sections for exclusive $J/\Psi$ and
  $\eta_c$ production}

The amplitude for exclusive production of a $J/\Psi$ meson is given by
\be \label{eq:ATL_JPsi_P}
{\cal A}_{T,L}^{\gamma^*p\rightarrow J/\Psi p} (Q^2,K_T)= i \int
\dd^2r \int\limits_0^1 \frac{dz}{4\pi} \, \left(\Psi_{\gamma^*}
\Psi^*_{J/\Psi}\right)_{T,L}(\vec r,z,Q^2)~e^{-i\frac{(1-2z)}{2}\vec r\cdot
  \vec K_T} \,\, {\cal T}_{gg}(\vec r,\vec K_\perp)\,.
\ee
This is independent of the direction of the transverse momentum
transfer $\vec K_T$ and satisfies
\be
\left[{\cal A}_{\lambda}^{\gamma^*p\rightarrow J/\Psi p}
  (Q^2,\vec K_T)\right]^* = -
{\cal A}_{\lambda}^{\gamma^*p\rightarrow J/\Psi p} (Q^2,-\vec K_T) = -
{\cal A}_{-\lambda}^{\gamma^*p\rightarrow J/\Psi p} (Q^2,-\vec K_T)~,
\ee
since it does not depend on the sign of $\lambda$ either. This last
expression verifies the analyticity property of the S-matrix, $S= 1 +
{\cal A}_\lambda(\vec K_T)$, for elastic scattering.

In \eq{eq:ATL_JPsi_P},  $\Psi_{J/\Psi}$ and $\Psi_{\gamma^*}$ denote
the $J/\psi$ and virtual photon light cone wave functions (for
longitudinal or transverse polarization); their product is summed over
the helicities of the $c$ and $\bar c$ quarks. We use the expressions
given in ref.~\cite{KMW} for numerical estimates, see
appendix~\ref{sec:LCwf}.

${\cal T}_{gg}(\vec r,\vec K_\perp)$ is independent of $x$ due to us
neglecting QCD evolution. On physical grounds, the resulting amplitude
${\cal A}_{T,L}^{\gamma^*p\rightarrow J/\Psi p}$ should provide a
first approximation for a collision energy such that $x\sim
0.1$. Greater $x$ are not accessible due to our assumption of eikonal
scattering with negligible longitudinal momentum transfer. In fact,
kinematic finite $x$ corrections may be substantial even at $x\sim
0.1$ and should be accounted for in the future.  From
\be \label{eq:x_vs_W}
x= \frac{Q^2 + M^2_{J/\Psi}}{2p\cdot q}= \frac{Q^2 + M^2_{J/\Psi}}{2m_pE_{\gamma^*}}
\ee
we estimate that $E_{\gamma^*}\simeq 25 - 50$~GeV (on a fixed target),
or $W^2\simeq 50-100$~GeV$^2$.  For such energy, and $Q^2 \simle
1$~GeV$^2$, the minimal value of $|t|$ due to longitudinal momentum
exchange is
\be \label{eq:tmin}
-t_\text{min} =
\left[ m_p \frac{Q^2+M^2_{J/\Psi}}{W^2}\right]^2 =
\left[ m_p \,\frac{Q^2+M^2_{J/\Psi}}{2m_pE_{\gamma^*}-Q^2+m_p^2}\right]^2
\simle 0.05~\text{GeV}^2~.
\ee
Since we neglect longitudinal momentum transfer, we restrict to
$|t-t_\text{min}| \simge 0.1$~GeV$^2$, and our values for $t$ should
be understood as corresponding to $t-t_\text{min}$.

In the high energy limit the differential cross section is given by~\cite{KMW}
\be \label{eq:sigma_AA*}
\frac{\dd\sigma}{\dd t} = \frac{1}{16\pi} \sum_{T,L}
\left|\,{\cal A}_{T,L}^{\gamma^* p\rightarrow M\, p}\,
\right|^2 \, .
\ee
On the r.h.s.\ of this equation, the squared amplitude can be
evaluated for arbitrary direction of $\vec K_T$ as it is invariant
under rotations.

\begin{figure}[htb]
  \centering
  \includegraphics[width=0.45\textwidth]{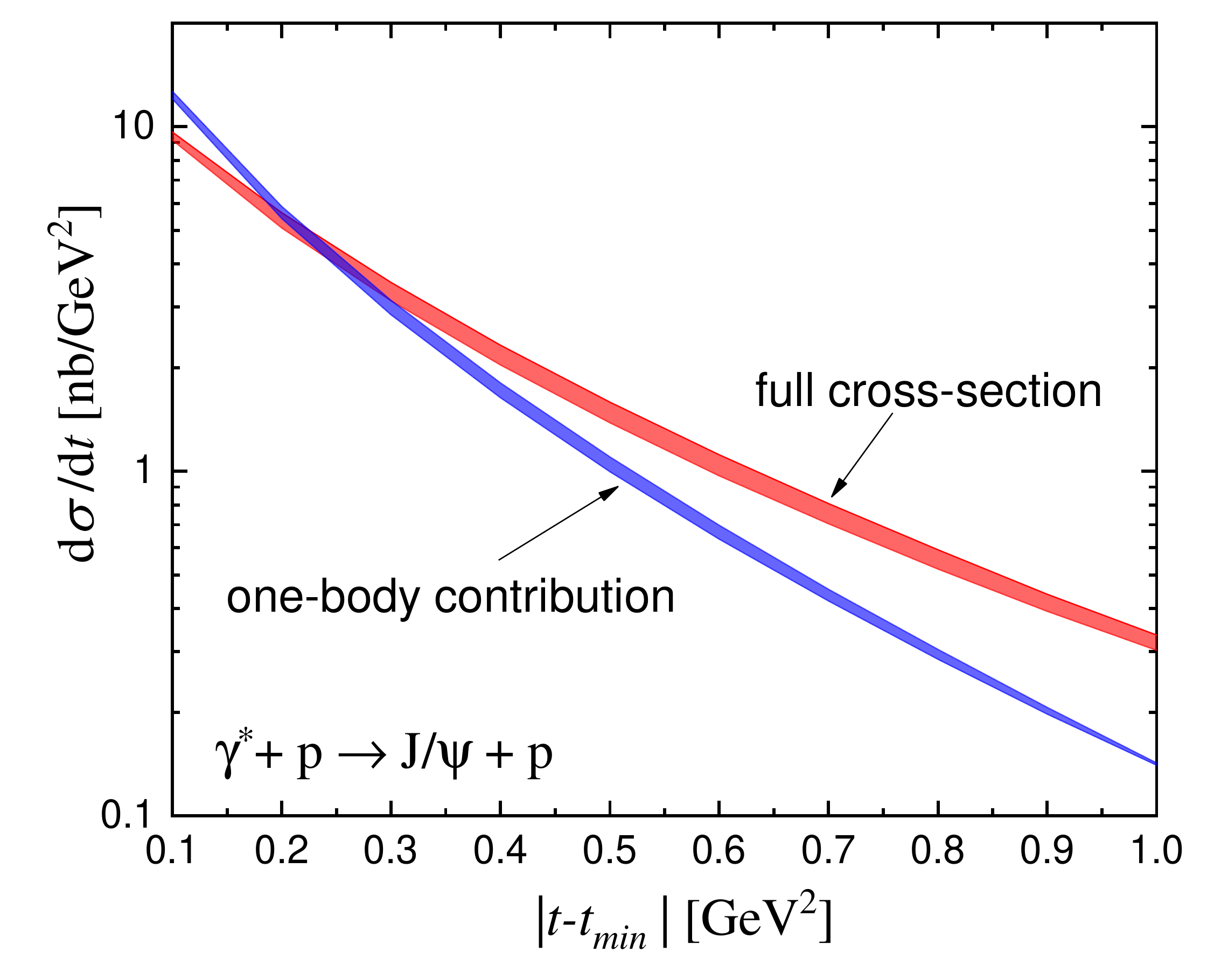}
  \caption{The differential cross section for exclusive $J/\Psi$
    production. The bands indicate the variation due to the two proton
    wave functions used here, and also cover the range $Q^2 <
    0.5$~GeV$^2$. The flatter curve is two gluon exchange
    with up to two quarks in the proton, while the steeper curve
    corresponds to the $c\bar c$ dipole scattering from a single quark
    in the proton.
    This figure is for a $\gamma^{(*)}-p$ collision energy of
    approximately $W\simeq 7-10$~GeV.}
\label{fig:dsigma_dt_JPsi}
\end{figure}
In \fig{fig:dsigma_dt_JPsi} we plot the resulting cross section for
$J/\Psi$ production. Fitting by an exponential fall-off over the range
$-0.5$~GeV$^2 > t-t_\text{min} > -1$~GeV$^2$ we obtain a slope of
$B\simeq3$~GeV$^{-2}$ which is close to data at comparable
energies~\cite{Camerini:1975cy}. On the other hand, a fit with a
(squared) dipole form factor~\cite{Frankfurt:2002ka}, $\dd\sigma/\dd t
\sim (1-(t-t_\text{min})/m_g^2)^{-4}$, can be performed over the
entire range $-0.1 > t -t_\text{min} > -1$~GeV$^2$. It results in
$m_g^2 \simeq 0.6$~GeV$^2$, which is close to $m_\rho^2$ but somewhat
smaller than $m_g^2 \simeq 1$~GeV$^2$ suggested in
ref.~\cite{Frankfurt:2002ka}.

Our expressions for the scattering amplitude have been obtained in a
fixed coupling approximation. Assuming $\alpha_s=0.35$, the integral
of $\dd\sigma/\dd t$ over $|t-t_\text{min}|>0.1$~GeV$^2$ is
$\sigma\approx 2.0$~nb; it increases by about a factor of 1.7 if we
extrapolate the integral all the way down to $t=t_\text{min}$.  Then
the differential cross section at $t=t_\text{min}$ is $\dd\sigma/\dd t
\approx 20$~nb/GeV$^2$.

The extrapolation to $t=t_\text{min}$ of course suffers from some
uncertainty due to neglecting longitudinal momentum transfer. Also,
there are uncertainties as to the values of $\alpha_s$ and
$m_Q$. Lastly, for more accurate results one should account for the
real part of the amplitude, too (see, for example,
refs~\cite{KMW,Forshaw:2003ki,Nemchik:1996cw,Martin:1999wb,Kharzeev:1998bz}).
However, $\dd\sigma/\dd t \approx 20$~nb/GeV$^2$ is not very far from
previous estimates for the $J/\Psi$ cross section at $W \simeq
7-10$~GeV~\cite{Kharzeev:1998bz}. The $J/\Psi$ cross section scales
with the coupling as $\sim \alpha_s^4$.

\fig{fig:dsigma_dt_JPsi} also shows the cross section obtained from
only the ``one body'' diagram where both exchanged gluons couple to
the same quark in the proton. This is a fair approximation to within a
factor of about 2 for $|t-t_\text{min}|\simle1$~GeV$^2$. The figure
illustrates the effect on the slope of $\dd \sigma/\dd t$ of the
diagrams where the $c\bar c$ dipole scatters from multiple quarks in
the proton\footnote{Charm production in photon scattering off multiple
  quarks has also been advocated in an unrelated setting for the
  $x\to1$ threshold region~\cite{Brodsky:2000zc}.}.  They lead to a
harder slope due to the contribution from configurations where the
exchanged gluons have similar transverse momenta. \\~~\\

The amplitude for elastic exclusive production of a $\eta_c$ meson is given by
\bea \label{eq:ATL_etac_iO}
{\cal A}_{\lambda}^{\gamma^*p\rightarrow \eta_c \,p} (Q^2,\vec K_T) &=& i \int
\dd^2r \int\limits_0^1 \frac{dz}{4\pi} \, \left(\Psi^{\gamma^*}_\lambda
(\Psi^{\eta_c})^*\right)(\vec r,z,Q^2)~e^{-i\frac{(1-2z)}{2}\vec r\cdot
  \vec K_T} \,\, \left({\cal T}_\gamma + {\cal T}_{\gamma gg} + {\cal T}_{ggg}
\right)(\vec r,\vec K_\perp)~.
\eea
Here, $\lambda=\pm$ denotes the polarization of the transverse photon,
the longitudinal photon does not contribute.

The product $\Psi^{\gamma^*}_\lambda
(\Psi^{\eta_c})^*$ changes sign under $\vec r\to -\vec r$ and therefore
\be
   {\cal A}_{\lambda}^{\gamma^*p\rightarrow \eta_c p} (Q^2,\vec K_T)= -
   {\cal A}_{\lambda}^{\gamma^*p\rightarrow \eta_c p} (Q^2,-\vec K_T) ~.
\ee
The complex conjugate amplitude is
\bea \label{eq:ATLcc_etac_iO}
\left[{\cal A}_{\lambda}^{\gamma^*p\rightarrow \eta_c p} (Q^2,\vec
  K_T)\right]^* &=&
     {\cal A}_{-\lambda}^{\gamma^*p\rightarrow \eta_c p} (Q^2,-\vec K_T)
\,.
\eea
This confirms that the S-matrix, $S(\vec K_T, \lambda) = 1+{\cal
  A}(\vec K_T, \lambda)$, exhibits the proper analytical properties
for elastic scattering.  We note that $\sum_\lambda |{\cal
  A}_{\lambda}^{\gamma^*p\rightarrow \eta_c \, p}|^2$ is invariant
under rotations of $\vec K_T$.\\

\begin{figure}[htb]
  \centering
  \includegraphics[width=0.48\textwidth]{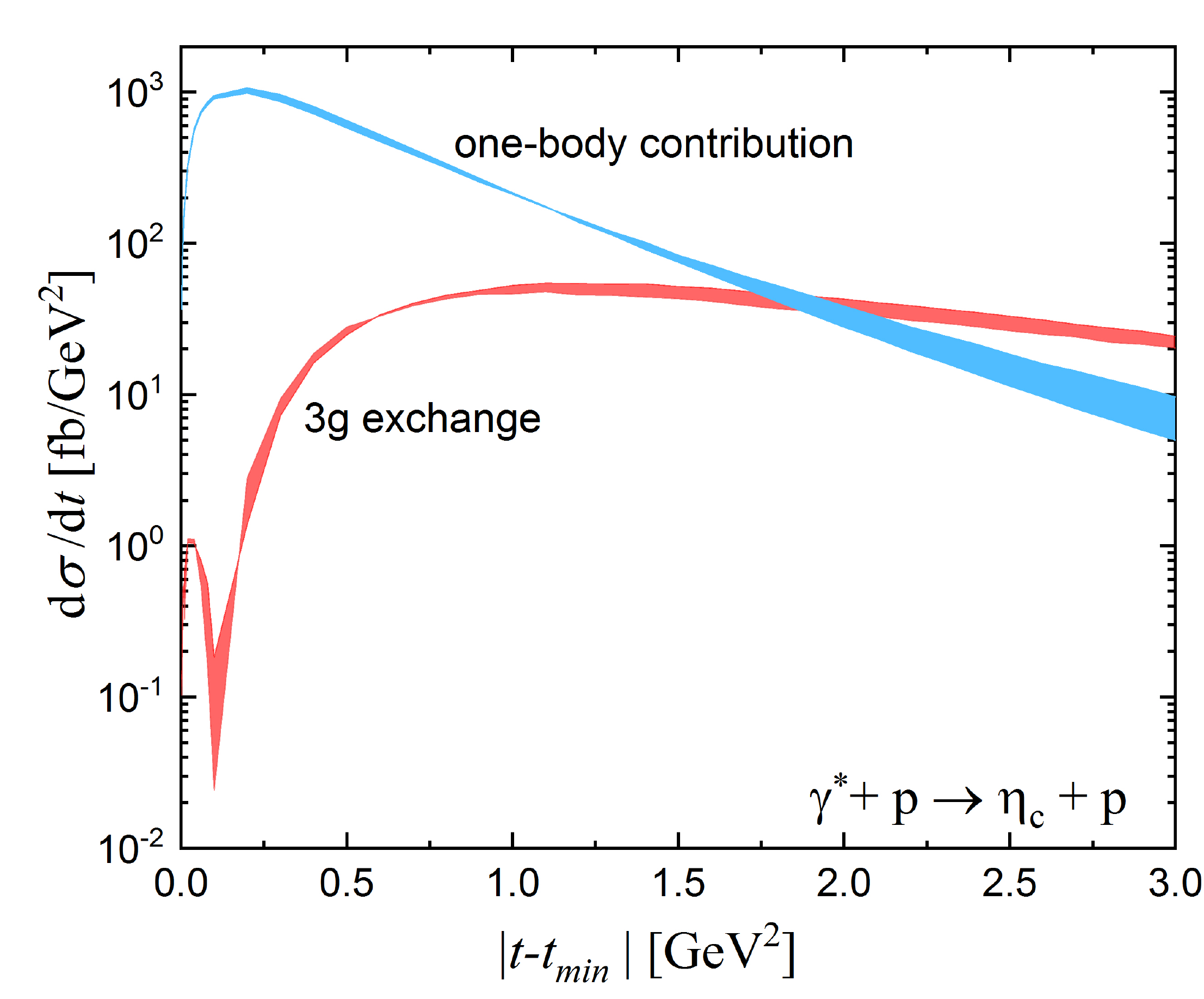}
  \includegraphics[width=0.48\textwidth]{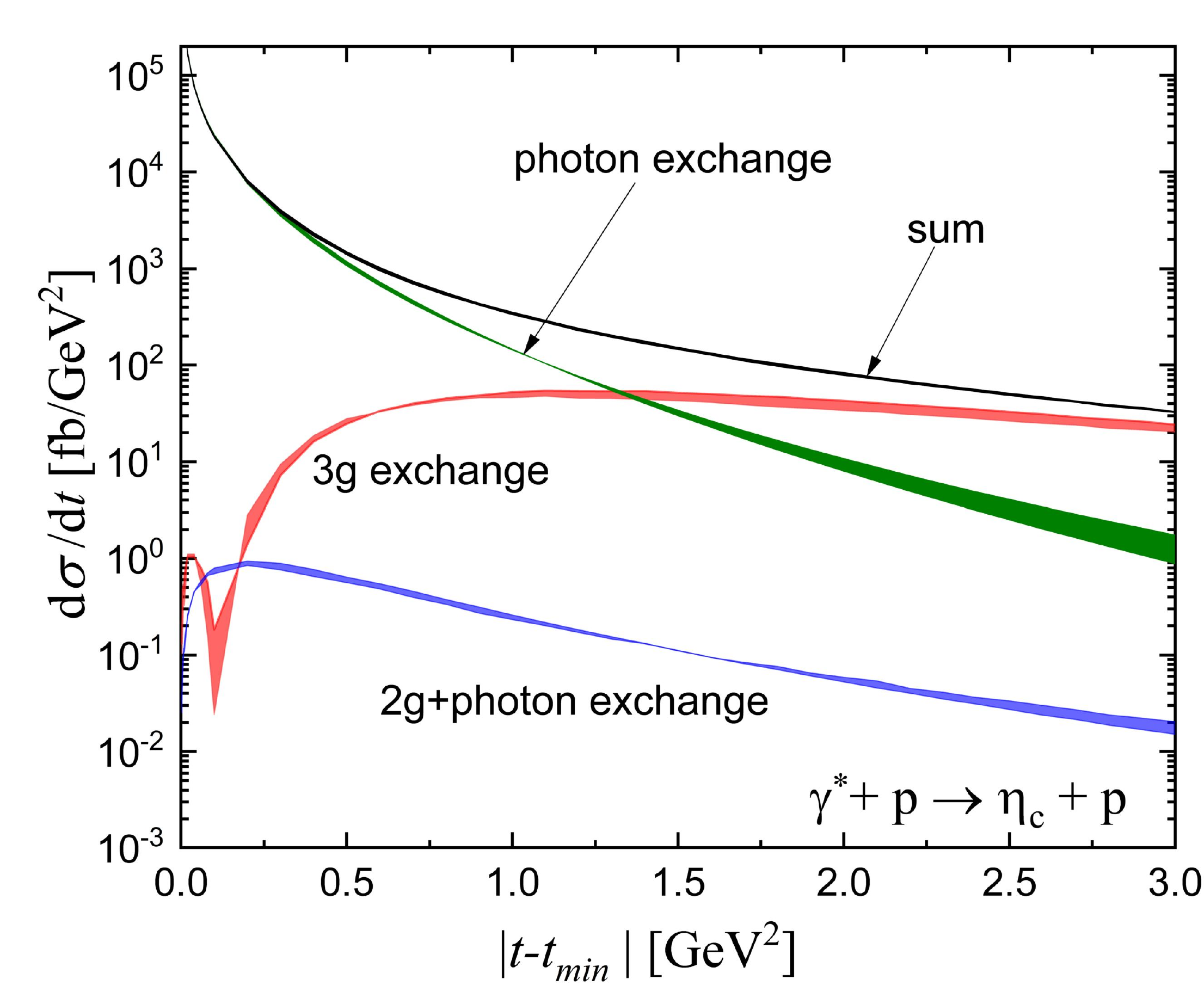}
  \caption{The differential cross section for exclusive $\eta_c$
    production. The bands indicate the variation due to the two proton
    wave functions used here, and also cover the range $Q^2 <
    0.5$~GeV$^2$.  This figure is for a $\gamma^{(*)}-p$ collision
    energy of approximately $W\simeq 7-10$~GeV. Left: the cross
    section due to three gluon exchange alone. The flatter curve
    accounts for all diagrams while the steeper curve corresponds to
    the $c\bar c$ dipole scattering from a single quark in the proton.
    Right: individual contributions due to single photon, photon plus
    two gluon, and three gluon exchanges, and the complete $\eta_c$
    cross section summed (at the amplitude level) over all these exchanges.}
\label{fig:dsigma_dt_etac}
\end{figure}
In fig.~\ref{fig:dsigma_dt_etac} we present our results for exclusive
production of $\eta_c$ mesons. We extend all curves down to
$K_T=0$ in order to show that the differential cross sections
due to $\gamma gg$ or $ggg$ exchange vanish in this limit. However, as
mentioned above, we do neglect longitudinal momentum transfer and so
we do not expect that these curves are reliable for $t\simeq
t_\text{min}$.

The left panel of fig.~\ref{fig:dsigma_dt_etac} shows the cross
section due to three gluon exchange. We compare the full cross section
which includes matrix elements between two and three quark states to
the one-body approximation with ad hoc infrared
cutoff~(\ref{eq:cutoff}). They exhibit very different behavior for
$|t| \simle 1.5$~GeV$^2$. It is interesting to note that the sum of
all three-gluon-exchange diagrams achieves its maximum at rather hard
$|t| \simge 1$~GeV$^2$. The dominant contribution corresponds to
approximate sharing of the momentum transfer among the three gluons,
$\vec q_i \sim - \vec K_T/3$. At the same time, gluons with transverse
momentum less than $\Lambda_\text{eff}$ are screened so that the cross
section below $|t|\sim 1$~GeV$^2$ decreases with $|t|$.

On the other hand, the $\eta_c$ cross section at $|t| \simle
1.5$~GeV$^2$ is anyhow dominated by single photon exchange as seen in
fig.~\ref{fig:dsigma_dt_etac} on the right. This appears to present
the cleanest opportunity for measuring the quark GPD $f(K_T)$ in this
process. The photon plus two gluon exchange is negligible over the
entire range of $|t|$.  ``Odderon'' exchange dominates for $|t|\simge
1.5$~GeV$^2$. Hence, discovery of the three gluon QCD exchange in this
process requires measurements at fairly large momentum transfer. Since
the dominant dipole scale for charmonium production is about $r\sim
1$~GeV$^{-1}$, at large $K_T$ one requires the dipole scattering
amplitude to all orders in $\vec r\cdot\vec K_T$.  \\

Our predictions for the $\eta_c$ cross section due to three gluon
exchange in $\gamma-p$ scattering are substantially lower than earlier
estimates\footnote{Those papers aimed at much higher HERA
  energies. However, their numerical estimates did not account for
  small-$x$ evolution of the Odderon and exhibit no energy dependence
  (for $W$ well above
  threshold).}~\cite{Czyzewski:1996bv,Engel:1997cga}. Their cross
section peaks at $\approx 10-25$~pb/GeV$^2$ at $|t| \approx
0.5$~GeV$^2$. In contrast, we find that the cross section due to three
gluon exchange increases with the momentum transfer up until
$|t|\sim1$~GeV$^2$, where its magnitude is about 50~fb/GeV$^2$, and
then decreases rather slowly for $|t|$ up to 3~GeV$^2$. Aside from
using a different approach for the three gluon exchange amplitude
which here is related to the light cone wave function of the proton,
and perhaps more realistic models for the wave function of the
$\eta_c$, we note the following
differences. Refs.~\cite{Czyzewski:1996bv,Engel:1997cga} employ a
coupling of the exchanged gluons to the valence quarks of the proton
of $\alpha_s=1$. We use the same $\alpha_s=0.35$ obtained from the
cross section for $J/\Psi$ production for all gluon-quark
vertices\footnote{F.~E.~Low estimated more than 40 years ago that
  $\alpha_s\approx\frac{1}{3}$ for two-gluon ``Pomeron''
  exchange~\cite{Low:1975sv}, so our qualitative extraction of
  $\alpha_s$ from the expected $J/\Psi$ cross section at $x\sim0.1$ is
  not a new result.}. This amounts to a suppression factor of $0.35^3
= 0.04$.  Furthermore, we find that the ``non-relativistic
approximation'' which sets the momentum fractions of the charm
quarks in the $\eta_c$ to 1/2 overestimates the cross section by as
much as a factor of 4 (this agrees with the findings in
ref.~\cite{Brodsky:1994kf}). We also use a more up to date value of
$\Gamma(\eta_c \to \gamma\gamma) = 5$~keV~\cite{PDG} to normalize the
$\eta_c$ wave function, and this is smaller than the value used in
refs.~\cite{Czyzewski:1996bv,Engel:1997cga} by a factor of
1.4. Lastly, as the maximum of the differential cross section is
shifted to higher $|t|$ it is natural that its magnitude would be
substantially lower.

\section{Summary}

Exclusive $\eta_c$ production in $\gamma-p$ scattering could provide
clean evidence for the semi-hard Odderon in QCD. Moreover, this
process would also provide valuable insight into the light cone wave
function of the proton which determines the coupling to both
Pomeron and Odderon simultaneously~\cite{DMV}. Accordingly, we first
applied our approach to $J/\Psi$ production, and from the magnitude of
the cross section, $\sigma^{J/\Psi}\sim \alpha_s^4$, we determined the
effective quark-gluon coupling, $\alpha_s\simeq0.35$. We also obtain a
reasonable slope of $\dd\sigma^{\gamma^*p\rightarrow J/\Psi p}/\dd t$
without having to tune any parameters. Most importantly, we find that
it is a fair approximation to consider $J/\Psi$ production (up to
intermediate $|t|\simeq1$~GeV$^2$) as due to two gluon exchange with
single quarks in the proton. This involves the matrix element of a
product of two color charge densities between single quark states and
can be expressed in terms of a leading twist Generalized Parton
Distribution (GPD).

${\cal C}$-parity even states like the $\eta_c$ can be produced only
via ${\cal C}$-odd exchanges such as a single photon, a photon and two
gluons, three gluons etc. Exchange of a single photon dominates at low
transverse momentum $K_T$ while the contribution from the exchange of
a photon and two gluons is very small for any $K_T$.  Our analysis of
three gluon exchange indicates that diagrams involving coupling of the
three gluons to multiple quarks are important not only for the
cancellation of infrared divergences due to color neutrality of the
proton. Rather, such many-body contributions are also numerically very
important for $|t|\simle 1$~GeV$^2$ (however, single photon exchange
dominates over three gluon exchange in that regime) and for $|t|\gg
1$~GeV$^2$. At high momentum transfer $K_T^2=|t|$ one may not expand
the dipole scattering amplitude in powers of $\vec r\cdot \vec K_T$.

A rather interesting outcome of our numerical analysis using the
proton light cone wave functions by Brodsky and
Schlumpf~\cite{Brodsky:1994fz} is that the differential cross section
for $\gamma^{(*)}p\rightarrow \eta_c p$ via three gluon exchange
achieves its maximum for $|t|\simeq1-3$~GeV$^2$. This is remarkable
since older estimates from the
literature~\cite{Czyzewski:1996bv,Engel:1997cga} using proton impact
factors dominated by the scale $m^2_\rho$ located the peak at a much
lower $|t|\simeq0.5$~GeV$^2$. A cross section on the order of (at
least) tens to a hundred fb/GeV$^2$ at $|t|>1.5$~GeV$^2$ would
represent good evidence for ${\cal C}$-odd three gluon exchange. On
the other hand, the best opportunity to measure the quark GPD in this
process is through single photon exchange at $|t|\simle 1.5$~GeV$^2$.

We estimate the cross section for $\eta_c$ production at
$|t|=1.5-3$~GeV$^2$ to be 30-150~fb/GeV$^2$, for $\alpha_s=0.35$ and
photon-proton collision energy $W\sim 7 - 10$~GeV. Experimental
detection therefore requires very high luminosities. For other values
of the coupling, the $J/\Psi$ and $\eta_c$ cross sections scale like
$\alpha_s^4$ and $\alpha_s^6$, respectively.

\section*{Acknowledgements}

A.D.\ acknowledges support by the DOE Office of Nuclear Physics
through Grant No.\ DE-FG02-09ER41620; and from The City University of
New York through the PSC-CUNY Research grant 60262-0048.

T.S.\ would like to thank the Ministry of Science and Higher Education of
Poland for support in the form of the Mobility Plus grant as well as
Brookhaven National Laboratory for hospitality and support. T.S.\ is also
supported by the Polish National Science Center (NCN) grant
No.\ 2017/27/B/ST2/02755.

We thank R.~Venugopalan for stimulating discussions that initiated
this work and for useful comments on the manuscript. We also thank
K.~Golec-Biernat, L.~Motyka and M.~Praszalowicz for helpful comments.

Figure~\ref{fig:3gXchange} has been prepared with
Jaxodraw~\cite{jaxo}.

\appendix

\section{Photon and $J/\Psi$ light cone wave functions} \label{sec:LCwf}

This appendix presents the expressions for the light-cone wave
functions of the photon and $J/\Psi$ meson used here. These
have been discussed in many
papers~\cite{KMW,Frankfurt:2002ka,Dosch:1996ss,Forshaw:2003ki}; we
follow the approach of ref.~\cite{KMW}. (For more recent numerical
solutions of quarkonium light cone wave functions
see~\cite{Li:2017mlw}.)

The wave function of a longitudinally polarized virtual photon is
given by~\cite{KMW}
\begin{equation}
  \Psi^{\gamma^*}_{h\bar{h},\lambda=0}(\vec r,z,Q^2) =  e_f e \, 
  \delta_{h,-\bar h} \, 2Qz(1-z)\, \frac{K_0(\epsilon r)}{2\pi}~.
  \label{long_photon_wf}
\end{equation}
For transverse polarization,
\begin{equation}
  \Psi^{\gamma^*}_{h\bar{h},\lambda=\pm}(\vec r,z,Q^2) =
  \lambda\, e_Q e \, \sqrt{2}\,
  \left\{
  \mathrm{i}e^{\lambda\, \mathrm{i}\phi_r}[
    z\delta_{h,\lambda}\delta_{\bar h,-\lambda} - 
    (1-z)\delta_{h,-\lambda}\delta_{\bar h,\lambda}] \partial_r \, + \, 
  m_Q \delta_{h,\lambda}\delta_{\bar h,\lambda}
  \right\}\, \frac{K_0(\epsilon r)}{2\pi}~.
  \label{transv_photon_wf}
\end{equation}
In the above expressions $K_0$ is a modified Bessel function of the
second kind, $\phi_r$ is the azimuthal angle of the $\vec r$ vector,
and $\epsilon^2 \equiv z(1-z)Q^2+m_f^2$. Note that in
eqs.~(\ref{long_photon_wf}) and~(\ref{transv_photon_wf}) we do not
include a factor of $\sqrt{N_c}$ as in ref.~\cite{KMW} because we have
explicitly included this factor in our expression for the scattering
amplitude. Ref.~\cite{Forshaw:2003ki} also includes a factor of
$1/\sqrt{4\pi}$ in the photon (and meson) wave function which we write
explicitly in eqs.~(\ref{eq:ATL_JPsi_P}, \ref{eq:ATL_etac_iO}) for the
amplitudes.\\

For the vector meson we employ the following model wave
functions~\cite{KMW,Brodsky:1994kf,Dosch:1996ss,Forshaw:2003ki}:
\begin{equation}
  \Psi^V_{h\bar{h},\lambda=0}(r,z) = \,
  \delta_{h,-\bar h} \,
  \left[ M_V\,+ \, \frac{m_f^2 - \nabla_r^2}{M_Vz(1-z)}\,  
    \right] \, \phi^{V}_L(r,z),
  \label{long_V_wf}
\end{equation}
and
\begin{equation}
  \Psi^V_{h\bar{h},\lambda=\pm 1}(\vec r,z) =
  \pm\, \frac{\sqrt{2}}{z(1-z)} \, 
  \left\{
  \mathrm{i}e^{\pm \mathrm{i}\phi_r}[
    z\delta_{h,\pm}\delta_{\bar h,\mp} - 
    (1-z)\delta_{h,\mp}\delta_{\bar h,\pm}] \partial_r \, + \, 
  m_f \delta_{h,\pm}\delta_{\bar h,\pm}
  \right\}\, \phi^{V}_T(r,z).
  \label{trans_V_wf}
\end{equation}
Several phenomenological models of the scalar part are available
in the literature. Here, we use the ``Boosted Gaussian'' model
\cite{KMW,Forshaw:2003ki,Nemchik:1994fp,Nemchik:1996cw}
\begin{align}
  \phi_{T,L}^{V}(r,z) = \mathcal{N}_{T,L}\, z(1-z)
  \exp\left(-\frac{m_f^2 \mathcal{R}^2}{8z(1-z)} -
  \frac{2z(1-z)r^2}{\mathcal{R}^2} +
  \frac{m_f^2\mathcal{R}^2}{2}\right)~.
  \label{Boosted-Gaussian_wf_model}
\end{align}
The parameters have been obtained in ref.~\cite{KMW} from the
normalization condition and from the electronic decay width. They are
$M_{J/\Psi}=3.097$~GeV, $m_c=1.4$~GeV, $\mathcal{N}_T=0.578$,
$\mathcal{N}_L=0.575$, $\mathcal{R}^2=2.3$~GeV$^{-2}$.\\

The product of photon and meson wave functions summed over quark
helicities, are given by
\be
\left[\left(\Psi^M\right)^*_\lambda \Psi^{\gamma^*}_{\lambda'}
  \right](\vec r,z,Q^2) \equiv
\sum_{\substack{h,\bar{h}=\pm}} \Psi_{h\bar{h},\lambda}^{M}(\vec
r,z)^*\, \Psi^{\gamma^*}_{h\bar{h},\lambda'}(\vec r,z,Q^2) ~.
\label{eq:overgg}
\ee
For a $J/\Psi$ meson, after averaging over $\lambda = \lambda' =+$ and
$\lambda = \lambda' =-$, one obtains explicitly~\cite{KMW}
\begin{align}
  \left[\left(\Psi^{J/\psi}\right)^* \Psi^{\gamma^*} \right]_T(\vec r,z,Q^2) &=
  e_c\, e\, \frac{1}{\pi z(1-z)} \,
  \left\{m_c^2 K_0(\epsilon r)\phi^{V}_T(r,z) -
  \left[z^2+(1-z)^2\right]\epsilon K_1(\epsilon r) \partial_r
  \phi^{V}_T(r,z)\right\},   \label{eq:overt}  \\
  \left[\left(\Psi^{J/\psi}\right)^* \Psi^{\gamma^*} \right]_L(\vec r,z,Q^2) &=
  \, e_c\, e \, \frac{1}{\pi}\, 2Qz(1-z)\,K_0(\epsilon r)\,
  \left[M_{J/\psi}\, \phi^{V}_L(r,z)+ \,\frac{m_c^2 -
      \nabla_r^2}{M_{J/\psi}\, z(1-z)} \phi^{V}_L(r,z)\right]~.
  \label{eq:overl}
\end{align}
In eq.~(\ref{eq:overt}) the polarizations $\lambda=\lambda'$ of the
photon and the $J/\Psi$ are equal; the off-diagonal overlap for
$\lambda=-\lambda'$ is proportional to $e^{2i\phi_r}$ and gives no
contribution to the amplitude~(\ref{eq:ATL_JPsi_P}) for $J/\Psi$
production as ${\cal T}_{gg}(\vec r,\vec K_T)$ is invariant under a
simultaneous rotation of $\vec r$ and $\vec K_T$. Also, the overlaps
$\left(\Psi^{J/\psi}_L\right)^* \Psi^{\gamma^*}_T$ and
$\left(\Psi^{J/\psi}_T\right)^* \Psi^{\gamma^*}_L$ do not vanish
either but change sign under $\vec r \to - \vec r$ and so do not
contribute to eq.~(\ref{eq:ATL_JPsi_P}).


\end{document}